\documentclass[aps,prl,twocolumn,nofootinbib,floatfix,superscriptaddress]{revtex4-2}
\usepackage{amsmath,amssymb,mleftright,mathtools}
\usepackage{microtype}
\usepackage{stix}
\usepackage{bm}
\usepackage{graphicx}
\graphicspath{{Figs/}}
\usepackage[colorlinks,linkcolor=blue,citecolor=blue,urlcolor=blue]{hyperref}
\usepackage[dvipsnames]{xcolor}

\def\a{\alpha}
\def\b{\beta}

\def\g{\gamma}

\def\e{\epsilon}
\def\be{\bar{\epsilon}}

\def\bm{\bar{\mu}}

\def\eps{\epsilon}
\def\s{\sigma}
\def\m{\mu}

\def\w{\omega}

\newcommand{\cA}{\mathcal{A}}
\newcommand{\cB}{\mathcal{B}}
\newcommand{\cC}{\mathcal{C}}
\newcommand{\cD}{\mathcal{D}}
\newcommand{\cE}{\mathcal{E}}

\def\bS{\bar{\Sigma}}

\def\bbS{\bar{\bar{\Sigma}}}

\def\G{\Gamma}
\def\S{\Sigma}
\def\W{\Omega}
\def\im{{\rm i}}
\def\ex{{\rm e}}

\def\nn{\nonumber}
\def\ud{\mathrm{d}}
\def\dd{\mathrm{d}}

\newcommand{\Ar}{\mathrm{A}}
\newcommand{\Rr}{\mathrm{R}}

\DeclareMathOperator{\re}{Re}
\DeclareMathOperator{\tr}{Tr}
\newcommand{\hc}{\hat{c}}

\allowdisplaybreaks
\begin{document}
%\title{Time-linear iterated generalized Kadanoff-Baym ansatz beyond wide-band approximation}
\title{Open system dynamics in linear-time beyond the wide-band limit}
\author{Y. Pavlyukh}
\affiliation{Institute of Theoretical Physics, Faculty of Fundamental Problems of Technology, Wroclaw University of Science and Technology, 50-370 Wroclaw, Poland}
\email{yaroslav.pavlyukh@pwr.edu.pl}
\author{R. Tuovinen}
\affiliation{Department of Physics, Nanoscience Center, P.O. Box 35, 40014 University of Jyv{\"a}skyl{\"a}, Finland}
\email{riku.m.s.tuovinen@jyu.fi}
\begin{abstract}
Nonequilibrium heat transport in quantum systems coupled to wide-band embeddings provides a striking example of the limitations of the generalized Kadanoff-Baym ansatz (GKBA), while solving the full two-time Kadanoff-Baym equations remains computationally prohibitive. To address this challenge, we propose an iterated solution to the reconstruction problem, resulting in a time-linear evolution scheme involving 14 correlators for systems with narrow-band embeddings. This approach eliminates GKBA-related artifacts and resolves convergence issues associated with the wide-band limit. Furthermore, it enables the calculation of energy- and time-resolved currents, facilitating the modeling of heat flows in quantum systems and energy- and time-resolved photoemission experiments, all at significantly reduced computational cost.
\end{abstract}
\maketitle

{\em Introduction.---}Thermodynamics predates electrodynamics, especially when considering foundational theories.
Because transport of charge plays such a fundamental technological role, the phenomenon is well understood and
formalized down to quantum, topological and nonequilibrium levels~\cite{meir_landauer_1992, chernodub_thermal_2022,
  kurth_time-dependent_2005}. On the other hand, thermal management in electronics and, in perspective, heat control in quantum processors,  nanoscale energy harvesting, quantum heat engines require similar understanding of the transport of
energy~\cite{dubi_colloquium_2011, pekola_colloquium_2021, arrachea_energy_2023}. This is closely related (Wiedemann-Franz law) but much less understood phenomenon:
Unlike electron current, heat flux is difficult to measure on the nanoscale~\cite{reddy_thermoelectricity_2007}. Since
multiple carriers and mechanisms are involved, it is difficult to formalize~\cite{luttinger_theory_1964, galperin_heat_2007}.

Further indications that heat flow is in many ways fundamentally different from charge flow arise from the consideration
of electron transport within the formalism of nonequilibrium Green's functions (NEGF)~\cite{stefanucci_nonequilibrium_2013, covito_transient_2018, ridley_many-body_2022}. In the time-linear formulation~\cite{tuovinen_time-linear_2023} derived utilizing the generalized Kadanoff-Baym ansatz (GKBA), the open system dynamics is described in terms of an “embedding correlator” from which the time-dependent current can be calculated using the Meir-Wingreen formula. The so-called wide-band limit approximation (WBLA) is a crucial ingredient of the theory allowing one to close the equations of motion. However, due to diverging energy integrals in the wide-band limit approximation, a simple generalization of the formalism towards the computation of heat currents represents a conceptual problem.

Besides this mathematical aspect, the validity of WBLA relies on the uniformity of the density of electronic states of the leads, which should be verified for each scenario~\cite{verzijl_applicability_2013, latini_charge_2014}.  In addition to transport experiments, the concept of embedding self-energy arises in the treatment of any open system, for instance, in the description of electron photoemission~\cite{schuler_time-dependent_2016, perfetto_first-principles_2016}, where energy-selective detection of particles requires departing from WBLA.

Therefore in this work, we lay down a time-linear scheme for open quantum systems with Lorentzian tunneling rates~\cite{tang_full-counting_2014}
\begin{equation}\label{eq:lorentzian}
\Gamma_{\a,ij}(\w) = \sum_{k} T_{ik\a}T_{k\a j}\delta(E_{k\a}-\w)=\frac{\gamma_{\a,ij}\Omega_\a^2}{(\w-\e_\a)^2+\Omega_\a^2},
\end{equation}
where $\e_\a$ and $\Omega_\a$ is the energy centroid and bandwidth, respectively, whereas $\gamma_{\a}$ is the level-width matrix depending on the coupling $T_{ik\a}$ between the state $i$ of central region to the state $k$ of the lead $\a$ with energy $E_{k\a}$. 

\paragraph{Density and currents.---}We focus here on the electronic transport mechanism and therefore start from the Kadanoff-Baym equation (KBE) for the electronic Green's function (GF) on the Keldysh contour $\mathcal{C}$,  
\begin{multline}\label{eq:gf-eom}
  \left[i \frac{d}{d z} - h(z) \right]G(z,z') = \delta(z,z') \\
  + \int_{\mathcal{C}} d\bar{z} \left[\Sigma_c(z,\bar{z})+\Sigma_{\mathrm{em}}(z,\bar{z})\right]G(\bar{z},z'),
\end{multline}
where $h$ is the one-electron mean-field (MF) Hamiltonian, $\Sigma_c$ is the correlation self-energy, and $\Sigma_{\mathrm{em}}=\sum_\a \Sigma_\a$ is the embedding self-energy (SE) due to leads---they are matrices in a suitably chosen basis. For simplicity of presentation, we include only the latter part into consideration noting that it is possible to incorporate electronic and electron-boson correlations as demonstrated in Ref.~\cite{tuovinen_time-linear_2023}. The electronic GF gives access to
different physical observables including currents
\begin{align}\label{eq:meir-wingreen}
  J_\a^{\nu}(t) &=\frac{d}{d t} \left\langle\textstyle\sum_{k}E_{k\a}^\nu \hc_{k\a}^\dagger \hc_{k\a}\right\rangle=2 \re \tr[I_\a^\nu(t)].
\end{align}
Here, $\hc_{k\a}$ operator is associated with noninteracting electrons of the leads, and charge ($\nu=0$) and energy ($\nu=1$) currents are expressed in terms of the collision integrals
\begin{align}
  \label{eq:collision}
 I_\a^\nu(t)=\int dt'\big[\Sigma_\a^{\nu,<}(t,t')G^{\mathrm{A}}(t',t)+\Sigma_\a^{\nu,\mathrm{R}}(t,t')G^<(t',t)\big].
\end{align}
The charge part was derived by Jauho, Meir and Wingreen~\cite{meir_landauer_1992, jauho_time-dependent_1994}, and later it was generalized to energy currents by several authors (see Ref.~\cite{kara_slimane_simulating_2020} and references therein). For brevity of notations we denote the ordinary self-energy as $\Sigma_\a^{0}=\S_\a$, and $\Sigma_\a^{1}$ entering the expression for the energy current will be defined below. In Eq.~\eqref{eq:meir-wingreen}, GFs and SEs are functions of real times, therefore lesser ($<$), retarded ($R$), and advanced ($A$) components appear by virtue of the Langreth rules~\cite{stefanucci_nonequilibrium_2013}.

Dealing with the lesser GF component in Eq.~\eqref{eq:collision} is a fundamental problem of the NEGF formalism. Splitting any lesser correlator into terms of retarded and advanced types
\begin{align}
  F^<(t,t')&=\theta(t-t')F^<_{\Rr}(t,t')-\theta(t'-t)F^<_{\Ar}(t,t'), \label{eq:split}
\end{align}
where $\theta(\tau)$ is the Heaviside step function, $G^<_{\Ar}$ in Eq.~\eqref{eq:collision} fulfills the celebrated exact reconstruction equation~\cite{lipavsky_generalized_1986}:
\begin{align}
  G^<_{\Ar}(t, t') &= -\rho(t) G^{\Ar}(t,t')-\int \dd{y}\dd{x}\big\{ G^{\Rr}(t, y)\S^<(y, x)\nn\\
  &+G_{\Rr}^<(t, y) \S^{\Ar}(y, x)\big\}
  \theta(x-t)G^{\Ar}(x,t'),\label{eq:reconstruction}
\end{align}
with $x$, $y$ denoting intermediate times. Over years, only the first part of this equation\,---\,what is conventionally
known as GKBA\,---\,was considered leading to the impressive progress in the treatment of correlated
electronic~\cite{schlunzen_achieving_2020, pavlyukh_photoinduced_2021},
bosonic~\cite{karlsson_fast_2021,pavlyukh_time-linear_2022-1}, and open~\cite{tuovinen_time-linear_2023, tuovinen_electroluminescence_2024} systems. The
key to this success was the ability to reformulate the complicated integro-differential equation~\eqref{eq:gf-eom}
into a system of ordinary differential equations (ODEs), which can conveniently be propagated with linear scaling in physical time. However, some limitations of the time-linear scheme have been identified such as unphysical electron
densities in electronic~\cite{joost_dynamically_2022} and electron-boson systems~\cite{pavlyukh_time-linear_2022}, as well as currents in open systems~\cite{kalvova_beyond_2019, kalvova_dynamical_2023}.

In a series of works~\cite{kalvova_beyond_2019, kalvova_dynamical_2023, kalvova_fast_2024}, Kalvov\'{a}, \v{S}pi\v{c}ka, Velick\'{y}, and Lipavsk\'{y} demonstrated dramatic improvements in the treatment of electronic transport, by approximately including correction terms in Eq.~\eqref{eq:reconstruction}. Their method involves the so-called Markovian~\cite{kalvova_dynamical_2023} or mirrored ($m$)~\cite{stefanucci_semiconductor_2024} GKBA $G_{\Rr}^<(x,z)=-\rho(x)G^{\Rr}(x, z)$ in the rhs of Eq.~\eqref{eq:reconstruction}. However, their hybrid scheme--—combining the conventional GKBA as a leading term with $m$GKBA in the corrections---relies on further approximations to evaluate the complex collision terms.

We demonstrate here that it is possible to deal with the reconstruction equation in a pure ODE way (akin of Ref.~\cite{tuovinen_time-linear_2023} for the wide-band limit) while retaining time-linear scaling and thus achieving high numerical efficiency. Crucially, no additional approximations are introduced beyond the use of GKBA $G_{\Rr}^<(x,z)=-G^{\Rr}(x, z)\rho(z)$ in the correction term of Eq.~\eqref{eq:reconstruction}, hence we term our method the iterated generalized Kadanoff-Baym ansatz ($i$GKBA). 

In the considered scenario, in addition to time-dependent MF $h(t)$, the dynamics is driven by voltages $V_\a(t)$ applied to the leads, and by the ramp function $s_\a(t)$ associated with tunneling matrix elements, i.e., $T_{ik\a}\rightarrow T_{ik\a} s_\a(t)$. This allows us to build a correlated initial state by the adiabatic switching procedure, and for the lead $\a$, results in the time-prefactor
\begin{align}
  s_\a(t) e^{-i\phi_\a(t,t')}s_\a(t')= s_\a(t)\s_\a(t,t')=s_\a(t) u_\a(t,t') s_\a(t'),
\end{align}
where $\phi_\a(t,t')\equiv \int_{t'}^t \dd{x} V_\a(x)$ is the accumulated phase due to the applied voltage. The SE components read~\cite{stefanucci_nonequilibrium_2013}:
\begin{subequations}
  \label{eq:sgm:emb}
\begin{align}
\Sigma_\a^{\nu, \mathrm{R}}(t,t') & = -i s_\a(t)\s_\a(t,t') \mathcal{F}\mleft[\w^\nu \Gamma_\a(\w)\mright](t-t')\theta(t-t'),\label{eq:sigmar}\\
\Sigma_\a^{\nu, <}(t,t') & = is_\a(t)\s_\a(t,t')\mathcal{F}\mleft[\w^\nu f_\a(\w)\Gamma_\a (\w)\mright](t-t'),\label{eq:sigmalss}
\end{align}
\end{subequations}
where $\mathcal{F}[a](\tau)=\int\frac{\ud \w}{2\pi}\ex^{-\im\w\tau}a(\w)$ is the Fourier transform and $f_\a(\w)$ is the Fermi-Dirac (FD) distribution function (for inverse temperature $\beta_\a$ and chemical potential $\mu_\a$), which we write using a suitable pole expansion~\cite{hu_communication:_2010}
\begin{multline}
  f_\a(\omega)=\frac{1}{e^{\beta_\a(\w-\mu_\a)}+1}=\frac12-\sum_{\ell\ge 1}\eta_\ell
  \Bigl[\frac{1}{\beta_\a(\w-\mu_\a)+i\zeta_\ell}\\+\frac{1}{\beta_\a(\w-\mu_\a)-i\zeta_\ell}\Bigr],\quad\text{with\; $\re \zeta_\ell>0$.}
  \label{eq:pole:exp}
\end{multline}
Performing the Fourier transforms~\eqref{eq:sgm:emb}  by closing the integration contour in the complex lower, upper half-plane we obtain for $\S_\a^{\Rr}(t,x)= s_\a(t)\bS_\a^{\Rr}(t,x)$,  $\S_\b^{\Ar}(x, t)=\bS_\b^{\Ar}(x, t)s_\b(t)$, respectively:
\begin{subequations}
\begin{align}
  \bS_\a^{\Rr}(t, x)&= -\frac{i}{2} \W_\a\gamma_\a u_\a(t, x)s_\a(x)e^{-i\be_\a (t-x)}\theta(t-x),\label{eq:S:R:fin}\\
  \bS_\b^{\Ar}(x, t)&=\frac{i}{2} \W_\b \gamma_\b e^{i\be_\b^*(t-x)}s_\b(x)u_\b(x, t) \theta(t-x).\label{eq:S:A:fin}
\end{align}
\end{subequations}
where we introduced $\be_\a=\e_\a-i\W_\a$ and used $x$ to denote intermediate times in convolution integrals, and Greek subscripts to enumerate leads ($1\le \a,\b\le N_{\text{leads}}$).
\begin{widetext}
For the lesser component of embedding SE [$\S_{\Rr,\a}^{<}(t,x)= s_\a(t)\bS_{\Rr,\a}^{<}(t,x)$ and $\S_{\Ar,\b}^{<}(x, t)=\bS_{\Ar,\b}^{<}(x, t)s_\b(t)$] we analogously have:
\begin{align}
  \bS_{\Rr,\a}^{<}(t, x)&=
i u_\a(t, x)\Bigg\{\gamma_\a\frac{\W_\a}{2} f_\a(\e_\a-i\W_\a)e^{-i\be_\a (t-x)}
+i \sum_{\ell\ge1} \frac{\eta_\ell}{\b_\a}\G_\a\mleft(\m_\a-i\frac{\zeta_\ell}{\b_\a}\mright)
e^{-i \left(\m_\a-i\frac{\zeta_\ell}{\b_\a} \right)(t-x)}\Bigg\}s_\a(x)\nn\\
&=u_\a(t,x) s_\a(x)\sum_{\ell\ge0}\bar{\eta}_{\a\ell}e^{-i\bar{\mu}_{\a\ell}(t-x)}
=\sum_{\ell\ge0}\bar{\eta}_{\a\ell}\bbS_{\Rr,\a\ell}^{<}(t, x). \label{eq:S:<:fin}
\end{align}
where the expansion coefficients $\bar{\eta}_{\a\ell}$ and the exponents $\bar{\mu}_{\a\ell}$ are given by
\begin{align}
  \bar{\eta}_{\a\ell}&=\begin{cases}
  i\frac{\gamma_\a}{2}\W_\a f_\a(\e_\a-i\W_\a)& \ell=0,\\
  -\frac{\eta_\ell}{\b_\a}\G_\a\mleft(\m_\a-i\frac{\zeta_\ell}{\b_\a}\mright)&\ell\ge 1;
  \end{cases}&
  \bar{\mu}_{\a\ell}&=
  \begin{cases}
  \e_\a-i\W_\a& \ell=0,\\
  \m_\a-i\frac{\zeta_\ell}{\b_\a}&\ell\ge 1.
  \end{cases}
\end{align}
Analogously, for the lesser SE of advanced type (remember the incorporated minus sign, cf. Eq.~\ref{eq:split}) we obtain:
\begin{align}
  \bS_{\Ar,\b}^{<}(x, t)&=u_\b(x, t) s_\b(x)\sum_{\ell\ge0}\bar{\eta}^*_{\b\ell}e^{i\bar{\mu}^*_{\b\ell}(t-x)}
=\sum_{\ell\ge0} \bar{\eta}^*_{\b\ell}\bbS_{\Ar,\b\ell}^{<}(x, t) .
\end{align}
\end{widetext}

Inserting the embedding SEs~\eqref{eq:S:R:fin} and \eqref{eq:S:<:fin} into Eq.~\eqref{eq:collision} yields four contributions to the collision integral. Defining
\begin{subequations}
\begin{align}
  [a f\cdot b](t,t')&=\smallint\! \dd{x}\,a(t,x)f(x)b(x,t'),\\
  [a f\cdot b]_{\Ar}(t,t')&=\theta(t'-t)[a f\cdot b](t,t')
\end{align}
\end{subequations}
to simplify notations, the first term in the collision integral~\eqref{eq:reconstruction} can be expressed in terms of
\begin{align}
  \cD_{\a\ell}^{c}(t)&=\bigl[\bbS_{\Rr,\a\ell}^<\cdot G^{\Ar}\bigr](t,t).  \label{def:Dc}
\end{align}
Second collision term in combination with the first term of the reconstruction equation leads to
\begin{align}
  \cD_{\a}^{d}(t)&=\mleft[ \bS^{\Rr}_{\a}\rho\cdot G^{\Ar}\mright](t,t). \label{def:Dd}
\end{align}
Finally, the correction terms in Eq.~\eqref{eq:reconstruction} give rise to 
\begin{subequations}
  \label{def:A}
\begin{align}
  \cA_{\a\b}^{a}(t)&=\mleft[\bS_\a^{\Rr}\cdot\big[G^{\Rr}\cdot\S^<_\b\big]_{\Ar}\cdot G^{\Ar}\mright](t, t),\\
  \cA_{\a\b}^{b}(t)&=\mleft[\bS_\a^{\Rr}\cdot \big[G^{\Rr}\rho\cdot\S^{\Ar}_\b\big]_{\Ar}\cdot G^{\Ar}\mright](t,t).
  \label{def:Aa}
\end{align}
\end{subequations}
In the hybrid scheme $i$GKBA(h), which arises from using $m$GKBA in the correction part of Eq.~\eqref{eq:reconstruction}, $\rho$ in Eq.~\eqref{def:Aa} is positioned after $G^{\Rr}$. This slightly modifies the equations of motion, without any pronounced physical effect, as numerically demonstrated below.

\begin{figure}
 \includegraphics[width=0.95\columnwidth]{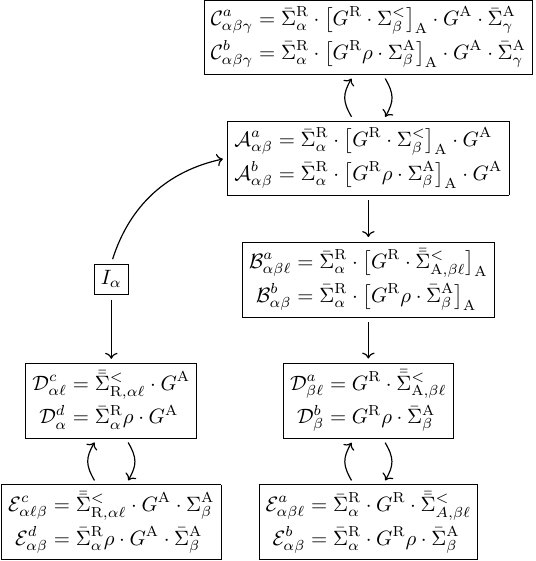}
\begin{caption}{\small
      Derivation of EOMs for the $i$GKBA correlators.    \label{fig:diagram}}
\end{caption}
\end{figure}

Due to the simple form of the embedding self-energy, it can be shown that the correlators satisfy a system of coupled ODEs. Consider for instance the time-derivative of $\cA^{a}(t)$:
  \begin{multline}
  i\frac{d}{dt} \cA_{\a\b}^{a}(t)=\cA_{\a\b}^{a}(t)\left[V_\a(t)+\e_\a-i\W_\a-h^\dagger(t)\right]\\
  -s_\b(t)\Bigl[\bS_\a^{\Rr}\cdot \bigl[G^{\Rr}\cdot \bS^<_\b\bigr]_{\Ar}\Bigr](t)\\
  -\sum_\g s_\g(t)\Bigl[\bS_\a^{\Rr}\cdot \bigl[G^{\Rr}\cdot \S^<_\b\bigr]_{\Ar}\cdot G^{\Ar}\cdot \bS^{\Ar}_\g\Bigr](t).
  \label{eom:A}
  \end{multline}
The first oscillatory term originates from the derivative of $u_\a$, the exponent in $\bS_\a^{\Rr}$~\eqref{eq:S:R:fin}, and from the mean-field part of the EOM for $G^{\Ar}(y,t)$. The simpler second term results from the differentiation of the $\theta$-function in $\bS_\a^{\Rr}$ and $G^{\Ar}$ yielding a $\delta$-function contribution, which can easily be integrated over. Finally, the third term results from the correlation term in the EOM for $G^{\Ar}(y,t)$. It leads to a more complicated correlator. The process of derivation is visualized in Fig.~\ref{fig:diagram}, where it is shown that $\cA_{\a\b}^{a}$ gives rise to $\cB_{\a\b\ell}^{a}$ and $\cC_{\a\b\g}^{a}$ correlators. Other correlators are treated similarly leading to in total 14 functions defined in Fig.~\ref{fig:diagram}. The corresponding EOMs follow a structure analogous to Eq.~\eqref{eom:A}; while straightforward, they are too lengthy to present here (see Supplemental Material~\cite{suppmat}). The overall computational scaling is $\mathcal{O}\bigl(N_\text{sys}^3 N_\text{leads}^2(2N_\text{leads}+4N_p)\bigr)$.

The collision integral~\eqref{eq:collision} can be written as:
\begin{align}
  I_\a(t)&=\bigl[\Sigma_\a^{<}\cdot G^{\Ar}+\Sigma_\a^{\Rr}\cdot G^<\bigr](t,t)\nn\\
  &=s_\a(t)\Bigl[\sum_{\ell\ge0}\bar{\eta}_{\a\ell} \bigl[\bbS_{\a\ell}^{<}\cdot G^{\Ar}\bigr]
    -\bigl[\bS_\a^{\Rr}\cdot G^<_{\Ar}\bigr]\Bigr](t,t) \label{eq:Ia:beyond:GKBA}\\
  &=s_\a(t)\Bigl\{\sum_{\ell\ge0} \bar{\eta}_{\a\ell} \cD_{\a\ell}^{c}(t)+\cD_\a^{d}(t)
    +\sum_\b \bigl[\cA_{\a\b}^a-\cA_{\a\b}^b\bigr](t)\Bigr\}.\nn 
\end{align}
To get the last equation, we substituted for $G^<_{\Ar}$ the reconstruction equation~\eqref{eq:reconstruction} and
used explicit definitions of $\cA$~\eqref{def:A} and $\cD$~(\ref{def:Dc}, \ref{def:Dd}) correlators.
Knowing $I_\a(t)$, density matrix can be propagated $\frac{d}{dt}\rho(t)=-\big(i h(t)\rho(t) +
\sum_\a I_\a(t)\big) + h.c.$,  and charge current can be computed according to Eq.~\eqref{eq:meir-wingreen}.

\begin{figure}
 \includegraphics[width=0.995\columnwidth]{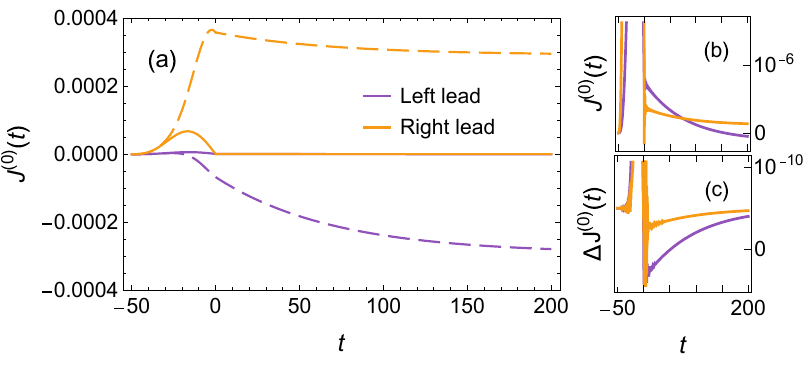}
\begin{caption}{\small
      (a) Electric currents computed within GKBA (dashed) and $i$GKBA (full) methods. (b) Deviation of the $i$GKBA results from zero. (c) Difference between the $i$GKBA(h) and $i$GKBA currents. High accuracy is ensured by the use of an adaptive step-size ODE solver.\label{fig:2}}
\end{caption}
\end{figure}

For the energy current, $I^{(1)}_\a(t)$ is needed. This collision integral is computed similarly to
Eq.~\eqref{eq:Ia:beyond:GKBA}, $I^{(1)}_\a(t)=\bigl[\Sigma_\a^{1,<}\cdot G^{\Ar}+\Sigma_\a^{1,\Rr}\cdot G^<\bigr](t,t)$.
Starting from the Fourier representations,  the relations between $\nu=0$ and $\nu=1$ self-energies follow $\bS_\a^{1,\Rr}=\be_\a\bS_\a^{\Rr}$, $\bS_{\Rr,\a}^{1,<}=\sum_{\ell\ge0}\bm_{\a\ell}\bar{\eta}_{\a\ell}\bbS_{\Rr,\a\ell}^{<}$
%\begin{align}
%  \bS_\a^{1,\Rr}&=\be_\a\bS_\a^{\Rr},&
%  \bS_{\Rr,\a}^{1,<}&=\sum_{\ell\ge0}\bm_{\a\ell}\bar{\eta}_{\a\ell}\bbS_{\Rr,\a\ell}^{<},
%\end{align}
leading to
\begin{multline}
  I^{(1)}_\a(t)=s_\a(t)\Bigl\{\sum_{\ell\ge0} \bm_{\a\ell}\bar{\eta}_{\a\ell} \cD_{\a\ell}^{c}(t) +\be_\a\cD_\a^{d}(t)\\
  +\be_\a\sum_\b \bigl[\cA_{\a\b}^a-\cA_{\a\b}^b\bigr](t)\Bigr\}.\label{eq:Ja:beyond:GKBA}
\end{multline}
Eqs.~(\ref{eq:Ia:beyond:GKBA} and \ref{eq:Ja:beyond:GKBA}), together with EOMs for the constituent $\cA$, $\cB$, $\cC$, $\cD$, and $\cE$ correlators represent the main result of this work. 
\paragraph{Numerical demonstration.---}A stringent test for open system calculations with narrow-band leads involves a scenario where a central system is connected to two leads with aligned chemical potentials and identical temperatures. From the Landauer–B\"{u}ttiker (LB) formula~\cite{stefanucci_nonequilibrium_2013}
\begin{multline}
  J_{\a}^{(0)}=\int\frac{d\w}{2\pi}\sum_\b\left(f_\a(\w-V_\a)-f_\b(\w-V_\b)\right)\\
  \times\tr\mleft[
    \G_\a(\w-V_\a) G^{\Rr}(\w)\G_\b(\w-V_\b) G^{\Ar}(\w)
    \mright]\label{eq:LB}
\end{multline}
such a system should exhibit no steady-state currents, even when the spectral densities $\G_\a$ and $\G_\b$ are centered at different energies~\cite{comment}. In our first setup, a central site with energy $\varepsilon_0=-3$ is connected to two leads with $\mu_1=\mu_2=0$, inverse temperatures $\b_1=\b_2=30$, and level widths $\g_1=\g_2=0.5$.  The initial state is prepared by adiabatically turning on the hopping to the leads on the time interval $[t_i,0]$, $t_i=-50$ using the ramp functions $s_{\a}(t)=\cos\mleft(\pi/2\cdot t/t_i\mright)^2\theta(-t)+\theta(t)$. For $t>0$, the system evolves without any applied voltages. In Fig.~\ref{fig:2}, we compare charge currents computed using different methods for the leads centered at $\eps_1=-10$ and $\eps_2=10$.

The standard GKBA approximation obtained by neglecting the $\cA$ correlators in Eq.~\eqref{eq:Ia:beyond:GKBA}, yields physically incorrect results, as indicated by the dashed lines. Persistent currents flow between the leads despite the absence of a driving force. Notably, there is no charge accumulation on the central site, as the sum of the currents remains zero.  The $i$GKBA and $i$GKBA(h) theories provide a much more accurate physical description: the steady-state currents remain around $10^{-6}$ (panel b), with their difference being even smaller, $\sim 10^{-10}$ (panel c).

\begin{figure}
 \includegraphics[width=0.995\columnwidth]{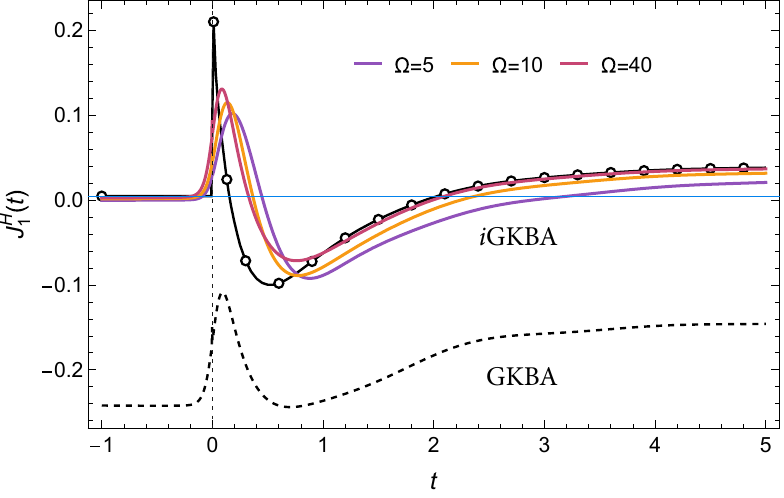}
\begin{caption}{\small
      Heat currents of a resonant level system. Dots represent exact analytic results for wide-band leads. $i$GKBA calculations for three different  $\W$ values show substantial improvement over the GKBA result ($\W=40$, dashed line). Blue line denotes a stationary LB value.\label{fig:3}}
\end{caption}
\end{figure}

Following this sanity check, we proceed to investigate more complex, bias-driven scenarios. We consider a resonant-level system described Kara Slimane \emph{et al}.~\cite{kara_slimane_simulating_2020}. In this \emph{noninteracting} model a central site with energy $\varepsilon_0(t)=0.5$ is connected to two leads ($\g_{1,2}=0.5$) with different temperatures and chemical potentials: $\b_1=1$, $\mu_1=0.5$ and $\b_2=10$, $\mu_2=-0.5$. The initial state is prepared by  following the same adiabatic protocol.  At $t=0$, time-dependent bias voltages are applied $V_\a(t)=-2.5[1+\exp(-25t)]^{-1}$.  In Fig.~\ref{fig:3} we focus on the heat currents defined as
\begin{align}
  J_\a^{\mathrm{H}}(t)&=J_\a^{(1)}(t)-\mu_\a J_\a^{(0)}(t).
\end{align}
Here, the GKBA (dashed line) exhibits a notable artifact: a significant negative shift in the heat current, attributable solely to inaccuracies in the energy current $J^{(1)}$, as all considered methods accurately predict the occupation of the central site $\rho(t)$ and the charge currents $J^{(0)}(t)$~\cite{suppmat}. Fig.~\ref{fig:3} demonstrates that $i$GKBA results converge toward the exact analytic WBLA solution as the spectral broadening $\W$ increases. Note that calculations for higher values of $\beta_\a$ and $\e_\a-i\W_\a$ require more terms ($N_p$) in the partial fraction expansion~\eqref{eq:pole:exp} of $f_\a(\be_\a)$.

\begin{figure}[b!]
 \includegraphics[width=0.995\columnwidth]{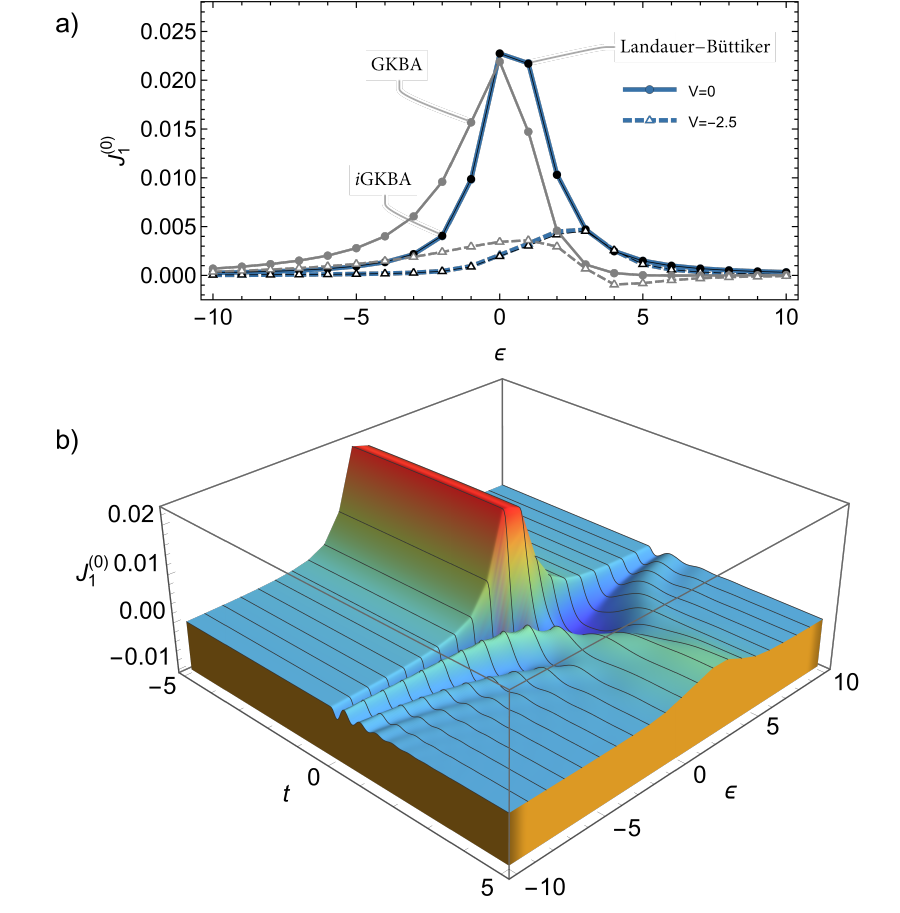}
\begin{caption}{\small
      Charge currents of a resonant-level system. Top: steady state currents before and after the voltage activation from the LB equation~\eqref{eq:LB} showing superiority of $i$GKBA over GKBA. Bottom: $i$GKBA energy- and time-resolved currents. \label{fig:4}}
\end{caption}
\end{figure}

Our approach is not restricted to Lorentzian spectral densities. By coupling the system to multiple leads, various shapes of the tunneling rates can be achieved, enabling energy- and time-resolved current calculations, Fig.~\ref{fig:4}. As an example, we extend the resonant-level model by replacing the left lead with 21 narrow sub-leads ($\W_{1j}=1$, $\b_{1j}=1$ and $\g_{1j}=0.16$) centered at $\eps_j=j$ with $-10\le j\le10$. The right lead parameters remain $\W_2=15$, $\b_2=10$ and $\g_2=0.5$. Simulations under the same excitation protocol illustrate the distinct spectral distribution of currents in the stationary states before and after bias activation, where $i$GKBA agrees perfectly with the LB prediction, along with a complex transient evolution.

\paragraph{Conclusion.---}
The achieved progress is therefore two-fold: i) We developed a time-linear nonequilibrium theory for the treatment of open systems coupled to \emph{narrow-band leads}. ii) As conventional GKBA is insufficient for this scenario, we put forward an $i$GKBA+ODE scheme corresponding to the first iteration of the reconstruction equation. Moreover, we resolved a debated issue about the choice of approximation for $G^{\Ar/\Rr}(t-t')$ in GKBA~\cite{latini_charge_2014,  schlunzen_nonequilibrium_2016, tuovinen_comparing_2020, tuovinen_electron_2021, stahl_memory_2022, makait_time-dependent_2023, reeves_unimportance_2023}. For open systems, we demonstrate that such approximations are unnecessary and can be fully incorporated into our formalism. This framework paves the way for exploring higher-order iterative reconstructions and potential extensions to correlated electronic and bosonic systems.  We anticipate applications in studying nonequilibrium heat flows~\cite{strasberg_quantum_2017, pekola_colloquium_2021, arrachea_energy_2023, portugal_heat_2024} and modeling energy- and time-resolved photoemission experiments~\cite{boschini_time-resolved_2024}\,---\,scenarios beyond the reach of conventional GKBA approaches.

\section{acknowledgement}
R.T. acknowledges the EffQSim project funded by the Jane and Aatos Erkko Foundation.

%\bibliography{SmallLibrary,SuppMat}

\begin{thebibliography}{45}%
\makeatletter
\providecommand \@ifxundefined [1]{%
 \@ifx{#1\undefined}
}%
\providecommand \@ifnum [1]{%
 \ifnum #1\expandafter \@firstoftwo
 \else \expandafter \@secondoftwo
 \fi
}%
\providecommand \@ifx [1]{%
 \ifx #1\expandafter \@firstoftwo
 \else \expandafter \@secondoftwo
 \fi
}%
\providecommand \natexlab [1]{#1}%
\providecommand \enquote  [1]{``#1''}%
\providecommand \bibnamefont  [1]{#1}%
\providecommand \bibfnamefont [1]{#1}%
\providecommand \citenamefont [1]{#1}%
\providecommand \href@noop [0]{\@secondoftwo}%
\providecommand \href [0]{\begingroup \@sanitize@url \@href}%
\providecommand \@href[1]{\@@startlink{#1}\@@href}%
\providecommand \@@href[1]{\endgroup#1\@@endlink}%
\providecommand \@sanitize@url [0]{\catcode `\\12\catcode `\$12\catcode
  `\&12\catcode `\#12\catcode `\^12\catcode `\_12\catcode `\%12\relax}%
\providecommand \@@startlink[1]{}%
\providecommand \@@endlink[0]{}%
\providecommand \url  [0]{\begingroup\@sanitize@url \@url }%
\providecommand \@url [1]{\endgroup\@href {#1}{\urlprefix }}%
\providecommand \urlprefix  [0]{URL }%
\providecommand \Eprint [0]{\href }%
\providecommand \doibase [0]{https://doi.org/}%
\providecommand \selectlanguage [0]{\@gobble}%
\providecommand \bibinfo  [0]{\@secondoftwo}%
\providecommand \bibfield  [0]{\@secondoftwo}%
\providecommand \translation [1]{[#1]}%
\providecommand \BibitemOpen [0]{}%
\providecommand \bibitemStop [0]{}%
\providecommand \bibitemNoStop [0]{.\EOS\space}%
\providecommand \EOS [0]{\spacefactor3000\relax}%
\providecommand \BibitemShut  [1]{\csname bibitem#1\endcsname}%
\let\auto@bib@innerbib\@empty
%</preamble>
\bibitem [{\citenamefont {Meir}\ and\ \citenamefont
  {Wingreen}(1992)}]{meir_landauer_1992}%
  \BibitemOpen
  \bibfield  {author} {\bibinfo {author} {\bibfnamefont {Y.}~\bibnamefont
  {Meir}}\ and\ \bibinfo {author} {\bibfnamefont {N.~S.}\ \bibnamefont
  {Wingreen}},\ }\bibfield  {title} {\bibinfo {title} {Landauer formula for the
  current through an interacting electron region},\ }\href
  {https://doi.org/10.1103/PhysRevLett.68.2512} {\bibfield  {journal} {\bibinfo
   {journal} {Phys. Rev. Lett.}\ }\textbf {\bibinfo {volume} {68}},\ \bibinfo
  {pages} {2512} (\bibinfo {year} {1992})}\BibitemShut {NoStop}%
\bibitem [{\citenamefont {Chernodub}\ \emph {et~al.}(2022)\citenamefont
  {Chernodub}, \citenamefont {Ferreiros}, \citenamefont {Grushin},
  \citenamefont {Landsteiner},\ and\ \citenamefont
  {Vozmediano}}]{chernodub_thermal_2022}%
  \BibitemOpen
  \bibfield  {author} {\bibinfo {author} {\bibfnamefont {M.~N.}\ \bibnamefont
  {Chernodub}}, \bibinfo {author} {\bibfnamefont {Y.}~\bibnamefont
  {Ferreiros}}, \bibinfo {author} {\bibfnamefont {A.~G.}\ \bibnamefont
  {Grushin}}, \bibinfo {author} {\bibfnamefont {K.}~\bibnamefont
  {Landsteiner}},\ and\ \bibinfo {author} {\bibfnamefont {M.~A.}\ \bibnamefont
  {Vozmediano}},\ }\bibfield  {title} {\bibinfo {title} {Thermal transport,
  geometry, and anomalies},\ }\href
  {https://doi.org/10.1016/j.physrep.2022.06.002} {\bibfield  {journal}
  {\bibinfo  {journal} {Phys. Rep.}\ }\textbf {\bibinfo {volume} {977}},\
  \bibinfo {pages} {1} (\bibinfo {year} {2022})}\BibitemShut {NoStop}%
\bibitem [{\citenamefont {Kurth}\ \emph {et~al.}(2005)\citenamefont {Kurth},
  \citenamefont {Stefanucci}, \citenamefont {Almbladh}, \citenamefont {Rubio},\
  and\ \citenamefont {Gross}}]{kurth_time-dependent_2005}%
  \BibitemOpen
  \bibfield  {author} {\bibinfo {author} {\bibfnamefont {S.}~\bibnamefont
  {Kurth}}, \bibinfo {author} {\bibfnamefont {G.}~\bibnamefont {Stefanucci}},
  \bibinfo {author} {\bibfnamefont {C.-O.}\ \bibnamefont {Almbladh}}, \bibinfo
  {author} {\bibfnamefont {A.}~\bibnamefont {Rubio}},\ and\ \bibinfo {author}
  {\bibfnamefont {E.~K.~U.}\ \bibnamefont {Gross}},\ }\bibfield  {title}
  {\bibinfo {title} {Time-dependent quantum transport: {A} practical scheme
  using density functional theory},\ }\href
  {https://doi.org/10.1103/PhysRevB.72.035308} {\bibfield  {journal} {\bibinfo
  {journal} {Phys. Rev. B}\ }\textbf {\bibinfo {volume} {72}},\ \bibinfo
  {pages} {035308} (\bibinfo {year} {2005})}\BibitemShut {NoStop}%
\bibitem [{\citenamefont {Dubi}\ and\ \citenamefont
  {Di Ventra}(2011)}]{dubi_colloquium_2011}%
  \BibitemOpen
  \bibfield  {author} {\bibinfo {author} {\bibfnamefont {Y.}~\bibnamefont
  {Dubi}}\ and\ \bibinfo {author} {\bibfnamefont {M.}~\bibnamefont
  {Di Ventra}},\ }\bibfield  {title} {\bibinfo {title} {\textit{{Colloquium}}
  : {Heat} flow and thermoelectricity in atomic and molecular junctions},\
  }\href {https://doi.org/10.1103/RevModPhys.83.131} {\bibfield  {journal}
  {\bibinfo  {journal} {Rev. Mod. Phys.}\ }\textbf {\bibinfo {volume} {83}},\
  \bibinfo {pages} {131} (\bibinfo {year} {2011})}\BibitemShut {NoStop}%
\bibitem [{\citenamefont {Pekola}\ and\ \citenamefont
  {Karimi}(2021)}]{pekola_colloquium_2021}%
  \BibitemOpen
  \bibfield  {author} {\bibinfo {author} {\bibfnamefont {J.~P.}\ \bibnamefont
  {Pekola}}\ and\ \bibinfo {author} {\bibfnamefont {B.}~\bibnamefont
  {Karimi}},\ }\bibfield  {title} {\bibinfo {title} {\textit{{Colloquium}} :
  {Quantum} heat transport in condensed matter systems},\ }\href
  {https://doi.org/10.1103/RevModPhys.93.041001} {\bibfield  {journal}
  {\bibinfo  {journal} {Rev. Mod. Phys.}\ }\textbf {\bibinfo {volume} {93}},\
  \bibinfo {pages} {041001} (\bibinfo {year} {2021})}\BibitemShut {NoStop}%
\bibitem [{\citenamefont {Arrachea}(2023)}]{arrachea_energy_2023}%
  \BibitemOpen
  \bibfield  {author} {\bibinfo {author} {\bibfnamefont {L.}~\bibnamefont
  {Arrachea}},\ }\bibfield  {title} {\bibinfo {title} {Energy dynamics, heat
  production and heat-work conversion with qubits: toward the development of
  quantum machines},\ }\href {https://doi.org/10.1088/1361-6633/acb06b}
  {\bibfield  {journal} {\bibinfo  {journal} {Rep. Prog. Phys.}\ }\textbf
  {\bibinfo {volume} {86}},\ \bibinfo {pages} {036501} (\bibinfo {year}
  {2023})}\BibitemShut {NoStop}%
\bibitem [{\citenamefont {Reddy}\ \emph {et~al.}(2007)\citenamefont {Reddy},
  \citenamefont {Jang}, \citenamefont {Segalman},\ and\ \citenamefont
  {Majumdar}}]{reddy_thermoelectricity_2007}%
  \BibitemOpen
  \bibfield  {author} {\bibinfo {author} {\bibfnamefont {P.}~\bibnamefont
  {Reddy}}, \bibinfo {author} {\bibfnamefont {S.-Y.}\ \bibnamefont {Jang}},
  \bibinfo {author} {\bibfnamefont {R.~A.}\ \bibnamefont {Segalman}},\ and\
  \bibinfo {author} {\bibfnamefont {A.}~\bibnamefont {Majumdar}},\ }\bibfield
  {title} {\bibinfo {title} {Thermoelectricity in {Molecular} {Junctions}},\
  }\href {https://doi.org/10.1126/science.1137149} {\bibfield  {journal}
  {\bibinfo  {journal} {Science}\ }\textbf {\bibinfo {volume} {315}},\ \bibinfo
  {pages} {1568} (\bibinfo {year} {2007})}\BibitemShut {NoStop}%
\bibitem [{\citenamefont {Luttinger}(1964)}]{luttinger_theory_1964}%
  \BibitemOpen
  \bibfield  {author} {\bibinfo {author} {\bibfnamefont {J.~M.}\ \bibnamefont
  {Luttinger}},\ }\bibfield  {title} {\bibinfo {title} {Theory of {Thermal}
  {Transport} {Coefficients}},\ }\href
  {https://doi.org/10.1103/PhysRev.135.A1505} {\bibfield  {journal} {\bibinfo
  {journal} {Phys. Rev.}\ }\textbf {\bibinfo {volume} {135}},\ \bibinfo {pages}
  {A1505} (\bibinfo {year} {1964})}\BibitemShut {NoStop}%
\bibitem [{\citenamefont {Galperin}\ \emph {et~al.}(2007)\citenamefont
  {Galperin}, \citenamefont {Nitzan},\ and\ \citenamefont
  {Ratner}}]{galperin_heat_2007}%
  \BibitemOpen
  \bibfield  {author} {\bibinfo {author} {\bibfnamefont {M.}~\bibnamefont
  {Galperin}}, \bibinfo {author} {\bibfnamefont {A.}~\bibnamefont {Nitzan}},\
  and\ \bibinfo {author} {\bibfnamefont {M.~A.}\ \bibnamefont {Ratner}},\
  }\bibfield  {title} {\bibinfo {title} {Heat conduction in molecular transport
  junctions},\ }\href {https://doi.org/10.1103/PhysRevB.75.155312} {\bibfield
  {journal} {\bibinfo  {journal} {Phys. Rev. B}\ }\textbf {\bibinfo {volume}
  {75}},\ \bibinfo {pages} {155312} (\bibinfo {year} {2007})}\BibitemShut
  {NoStop}%
\bibitem [{\citenamefont {Stefanucci}\ and\ \citenamefont {van
  Leeuwen}(2013)}]{stefanucci_nonequilibrium_2013}%
  \BibitemOpen
  \bibfield  {author} {\bibinfo {author} {\bibfnamefont {G.}~\bibnamefont
  {Stefanucci}}\ and\ \bibinfo {author} {\bibfnamefont {R.}~\bibnamefont {van
  Leeuwen}},\ }\href@noop {} {\emph {\bibinfo {title} {Nonequilibrium
  {Many}-{Body} {Theory} of {Quantum} {Systems}: {A} {Modern}
  {Introduction}}}}\ (\bibinfo  {publisher} {Cambridge University Press},\
  \bibinfo {address} {Cambridge},\ \bibinfo {year} {2013})\BibitemShut
  {NoStop}%
\bibitem [{\citenamefont {Covito}\ \emph {et~al.}(2018)\citenamefont {Covito},
  \citenamefont {Eich}, \citenamefont {Tuovinen}, \citenamefont {Sentef},\ and\
  \citenamefont {Rubio}}]{covito_transient_2018}%
  \BibitemOpen
  \bibfield  {author} {\bibinfo {author} {\bibfnamefont {F.}~\bibnamefont
  {Covito}}, \bibinfo {author} {\bibfnamefont {F.~G.}\ \bibnamefont {Eich}},
  \bibinfo {author} {\bibfnamefont {R.}~\bibnamefont {Tuovinen}}, \bibinfo
  {author} {\bibfnamefont {M.~A.}\ \bibnamefont {Sentef}},\ and\ \bibinfo
  {author} {\bibfnamefont {A.}~\bibnamefont {Rubio}},\ }\bibfield  {title}
  {\bibinfo {title} {Transient {Charge} and {Energy} {Flow} in the
  {Wide}-{Band} {Limit}},\ }\href {https://doi.org/10.1021/acs.jctc.8b00077}
  {\bibfield  {journal} {\bibinfo  {journal} {J. Chem. Theory Comput.}\
  }\textbf {\bibinfo {volume} {14}},\ \bibinfo {pages} {2495} (\bibinfo {year}
  {2018})}\BibitemShut {NoStop}%
\bibitem [{\citenamefont {Ridley}\ \emph {et~al.}(2022)\citenamefont {Ridley},
  \citenamefont {Talarico}, \citenamefont {Karlsson}, \citenamefont
  {Lo~Gullo},\ and\ \citenamefont {Tuovinen}}]{ridley_many-body_2022}%
  \BibitemOpen
  \bibfield  {author} {\bibinfo {author} {\bibfnamefont {M.}~\bibnamefont
  {Ridley}}, \bibinfo {author} {\bibfnamefont {N.~W.}\ \bibnamefont
  {Talarico}}, \bibinfo {author} {\bibfnamefont {D.}~\bibnamefont {Karlsson}},
  \bibinfo {author} {\bibfnamefont {N.}~\bibnamefont {Lo~Gullo}},\ and\
  \bibinfo {author} {\bibfnamefont {R.}~\bibnamefont {Tuovinen}},\ }\bibfield
  {title} {\bibinfo {title} {A many-body approach to transport in quantum
  systems: from the transient regime to the stationary state},\ }\href
  {https://doi.org/10.1088/1751-8121/ac7119} {\bibfield  {journal} {\bibinfo
  {journal} {J. Phys. A}\ }\textbf {\bibinfo {volume} {55}},\ \bibinfo {pages}
  {273001} (\bibinfo {year} {2022})}\BibitemShut {NoStop}%
\bibitem [{\citenamefont {Tuovinen}\ \emph {et~al.}(2023)\citenamefont
  {Tuovinen}, \citenamefont {Pavlyukh}, \citenamefont {Perfetto},\ and\
  \citenamefont {Stefanucci}}]{tuovinen_time-linear_2023}%
  \BibitemOpen
  \bibfield  {author} {\bibinfo {author} {\bibfnamefont {R.}~\bibnamefont
  {Tuovinen}}, \bibinfo {author} {\bibfnamefont {Y.}~\bibnamefont {Pavlyukh}},
  \bibinfo {author} {\bibfnamefont {E.}~\bibnamefont {Perfetto}},\ and\
  \bibinfo {author} {\bibfnamefont {G.}~\bibnamefont {Stefanucci}},\ }\bibfield
   {title} {\bibinfo {title} {Time-{Linear} {Quantum} {Transport} {Simulations}
  with {Correlated} {Nonequilibrium} {Green}'s {Functions}},\ }\href
  {https://doi.org/10.1103/PhysRevLett.130.246301} {\bibfield  {journal}
  {\bibinfo  {journal} {Phys. Rev. Lett.}\ }\textbf {\bibinfo {volume} {130}},\
  \bibinfo {pages} {246301} (\bibinfo {year} {2023})}\BibitemShut {NoStop}%
\bibitem [{\citenamefont {Verzijl}\ \emph {et~al.}(2013)\citenamefont
  {Verzijl}, \citenamefont {Seldenthuis},\ and\ \citenamefont
  {Thijssen}}]{verzijl_applicability_2013}%
  \BibitemOpen
  \bibfield  {author} {\bibinfo {author} {\bibfnamefont {C.~J.~O.}\
  \bibnamefont {Verzijl}}, \bibinfo {author} {\bibfnamefont {J.~S.}\
  \bibnamefont {Seldenthuis}},\ and\ \bibinfo {author} {\bibfnamefont {J.~M.}\
  \bibnamefont {Thijssen}},\ }\bibfield  {title} {\bibinfo {title}
  {Applicability of the wide-band limit in {DFT}-based molecular transport
  calculations},\ }\href {https://doi.org/10.1063/1.4793259} {\bibfield
  {journal} {\bibinfo  {journal} {J. Chem. Phys.}\ }\textbf {\bibinfo {volume}
  {138}},\ \bibinfo {pages} {094102} (\bibinfo {year} {2013})}\BibitemShut
  {NoStop}%
\bibitem [{\citenamefont {Latini}\ \emph {et~al.}(2014)\citenamefont {Latini},
  \citenamefont {Perfetto}, \citenamefont {Uimonen}, \citenamefont {van
  Leeuwen},\ and\ \citenamefont {Stefanucci}}]{latini_charge_2014}%
  \BibitemOpen
  \bibfield  {author} {\bibinfo {author} {\bibfnamefont {S.}~\bibnamefont
  {Latini}}, \bibinfo {author} {\bibfnamefont {E.}~\bibnamefont {Perfetto}},
  \bibinfo {author} {\bibfnamefont {A.-M.}\ \bibnamefont {Uimonen}}, \bibinfo
  {author} {\bibfnamefont {R.}~\bibnamefont {van Leeuwen}},\ and\ \bibinfo
  {author} {\bibfnamefont {G.}~\bibnamefont {Stefanucci}},\ }\bibfield  {title}
  {\bibinfo {title} {Charge dynamics in molecular junctions: {Nonequilibrium}
  {Green}'s function approach made fast},\ }\href
  {https://doi.org/10.1103/PhysRevB.89.075306} {\bibfield  {journal} {\bibinfo
  {journal} {Phys. Rev. B}\ }\textbf {\bibinfo {volume} {89}},\ \bibinfo
  {pages} {075306} (\bibinfo {year} {2014})}\BibitemShut {NoStop}%
\bibitem [{\citenamefont {Sch\"{u}ler}\ \emph {et~al.}(2016)\citenamefont
  {Sch\"{u}ler}, \citenamefont {Berakdar},\ and\ \citenamefont
  {Pavlyukh}}]{schuler_time-dependent_2016}%
  \BibitemOpen
  \bibfield  {author} {\bibinfo {author} {\bibfnamefont {M.}~\bibnamefont
  {Sch\"{u}ler}}, \bibinfo {author} {\bibfnamefont {J.}~\bibnamefont
  {Berakdar}},\ and\ \bibinfo {author} {\bibfnamefont {Y.}~\bibnamefont
  {Pavlyukh}},\ }\bibfield  {title} {\bibinfo {title} {Time-dependent many-body
  treatment of electron-boson dynamics: {Application} to plasmon-accompanied
  photoemission},\ }\href {https://doi.org/10.1103/PhysRevB.93.054303}
  {\bibfield  {journal} {\bibinfo  {journal} {Phys. Rev. B}\ }\textbf {\bibinfo
  {volume} {93}},\ \bibinfo {pages} {054303} (\bibinfo {year}
  {2016})}\BibitemShut {NoStop}%
\bibitem [{\citenamefont {Perfetto}\ \emph {et~al.}(2016)\citenamefont
  {Perfetto}, \citenamefont {Sangalli}, \citenamefont {Marini},\ and\
  \citenamefont {Stefanucci}}]{perfetto_first-principles_2016}%
  \BibitemOpen
  \bibfield  {author} {\bibinfo {author} {\bibfnamefont {E.}~\bibnamefont
  {Perfetto}}, \bibinfo {author} {\bibfnamefont {D.}~\bibnamefont {Sangalli}},
  \bibinfo {author} {\bibfnamefont {A.}~\bibnamefont {Marini}},\ and\ \bibinfo
  {author} {\bibfnamefont {G.}~\bibnamefont {Stefanucci}},\ }\bibfield  {title}
  {\bibinfo {title} {First-principles approach to excitons in time-resolved and
  angle-resolved photoemission spectra},\ }\href
  {https://doi.org/10.1103/PhysRevB.94.245303} {\bibfield  {journal} {\bibinfo
  {journal} {Phys. Rev. B}\ }\textbf {\bibinfo {volume} {94}},\ \bibinfo
  {pages} {245303} (\bibinfo {year} {2016})}\BibitemShut {NoStop}%
\bibitem [{\citenamefont {Tang}\ and\ \citenamefont
  {Wang}(2014)}]{tang_full-counting_2014}%
  \BibitemOpen
  \bibfield  {author} {\bibinfo {author} {\bibfnamefont {G.-M.}\ \bibnamefont
  {Tang}}\ and\ \bibinfo {author} {\bibfnamefont {J.}~\bibnamefont {Wang}},\
  }\bibfield  {title} {\bibinfo {title} {Full-counting statistics of charge and
  spin transport in the transient regime: {A} nonequilibrium {Green}'s function
  approach},\ }\href {https://doi.org/10.1103/PhysRevB.90.195422} {\bibfield
  {journal} {\bibinfo  {journal} {Phys. Rev. B}\ }\textbf {\bibinfo {volume}
  {90}},\ \bibinfo {pages} {195422} (\bibinfo {year} {2014})}\BibitemShut
  {NoStop}%
\bibitem [{\citenamefont {Jauho}\ \emph {et~al.}(1994)\citenamefont {Jauho},
  \citenamefont {Wingreen},\ and\ \citenamefont
  {Meir}}]{jauho_time-dependent_1994}%
  \BibitemOpen
  \bibfield  {author} {\bibinfo {author} {\bibfnamefont {A.-P.}\ \bibnamefont
  {Jauho}}, \bibinfo {author} {\bibfnamefont {N.~S.}\ \bibnamefont
  {Wingreen}},\ and\ \bibinfo {author} {\bibfnamefont {Y.}~\bibnamefont
  {Meir}},\ }\bibfield  {title} {\bibinfo {title} {Time-dependent transport in
  interacting and noninteracting resonant-tunneling systems},\ }\href
  {https://doi.org/10.1103/PhysRevB.50.5528} {\bibfield  {journal} {\bibinfo
  {journal} {Phys. Rev. B}\ }\textbf {\bibinfo {volume} {50}},\ \bibinfo
  {pages} {5528} (\bibinfo {year} {1994})}\BibitemShut {NoStop}%
\bibitem [{\citenamefont {Kara~Slimane}\ \emph {et~al.}(2020)\citenamefont
  {Kara~Slimane}, \citenamefont {Reck},\ and\ \citenamefont
  {Fleury}}]{kara_slimane_simulating_2020}%
  \BibitemOpen
  \bibfield  {author} {\bibinfo {author} {\bibfnamefont {A.}~\bibnamefont
  {Kara~Slimane}}, \bibinfo {author} {\bibfnamefont {P.}~\bibnamefont {Reck}},\
  and\ \bibinfo {author} {\bibfnamefont {G.}~\bibnamefont {Fleury}},\
  }\bibfield  {title} {\bibinfo {title} {Simulating time-dependent
  thermoelectric transport in quantum systems},\ }\href
  {https://doi.org/10.1103/PhysRevB.101.235413} {\bibfield  {journal} {\bibinfo
   {journal} {Phys. Rev. B}\ }\textbf {\bibinfo {volume} {101}},\ \bibinfo
  {pages} {235413} (\bibinfo {year} {2020})}\BibitemShut {NoStop}%
\bibitem [{\citenamefont {Lipavský}\ \emph {et~al.}(1986)\citenamefont
  {Lipavský}, \citenamefont {\v{S}pi\v{c}ka},\ and\ \citenamefont
  {Velický}}]{lipavsky_generalized_1986}%
  \BibitemOpen
  \bibfield  {author} {\bibinfo {author} {\bibfnamefont {P.}~\bibnamefont
  {Lipavský}}, \bibinfo {author} {\bibfnamefont {V.}~\bibnamefont
  {\v{S}pi\v{c}ka}},\ and\ \bibinfo {author} {\bibfnamefont {B.}~\bibnamefont
  {Velický}},\ }\bibfield  {title} {\bibinfo {title} {Generalized
  {Kadanoff}-{Baym} ansatz for deriving quantum transport equations},\ }\href
  {https://doi.org/10.1103/PhysRevB.34.6933} {\bibfield  {journal} {\bibinfo
  {journal} {Phys. Rev. B}\ }\textbf {\bibinfo {volume} {34}},\ \bibinfo
  {pages} {6933} (\bibinfo {year} {1986})}\BibitemShut {NoStop}%
\bibitem [{\citenamefont {Schl\"{u}nzen}\ \emph {et~al.}(2020)\citenamefont
  {Schl\"{u}nzen}, \citenamefont {Joost},\ and\ \citenamefont
  {Bonitz}}]{schlunzen_achieving_2020}%
  \BibitemOpen
  \bibfield  {author} {\bibinfo {author} {\bibfnamefont {N.}~\bibnamefont
  {Schl\"{u}nzen}}, \bibinfo {author} {\bibfnamefont {J.-P.}\ \bibnamefont
  {Joost}},\ and\ \bibinfo {author} {\bibfnamefont {M.}~\bibnamefont
  {Bonitz}},\ }\bibfield  {title} {\bibinfo {title} {Achieving the {Scaling}
  {Limit} for {Nonequilibrium} {Green} {Functions} {Simulations}},\ }\href
  {https://doi.org/10.1103/PhysRevLett.124.076601} {\bibfield  {journal}
  {\bibinfo  {journal} {Phys. Rev. Lett.}\ }\textbf {\bibinfo {volume} {124}},\
  \bibinfo {pages} {076601} (\bibinfo {year} {2020})}\BibitemShut {NoStop}%
\bibitem [{\citenamefont {Pavlyukh}\ \emph {et~al.}(2021)\citenamefont
  {Pavlyukh}, \citenamefont {Perfetto},\ and\ \citenamefont
  {Stefanucci}}]{pavlyukh_photoinduced_2021}%
  \BibitemOpen
  \bibfield  {author} {\bibinfo {author} {\bibfnamefont {Y.}~\bibnamefont
  {Pavlyukh}}, \bibinfo {author} {\bibfnamefont {E.}~\bibnamefont {Perfetto}},\
  and\ \bibinfo {author} {\bibfnamefont {G.}~\bibnamefont {Stefanucci}},\
  }\bibfield  {title} {\bibinfo {title} {Photoinduced dynamics of organic
  molecules using nonequilibrium {Green}'s functions with second-{Born},
  \textit{{GW}}, \textit{{T}}-matrix, and three-particle correlations},\ }\href
  {https://doi.org/10.1103/PhysRevB.104.035124} {\bibfield  {journal} {\bibinfo
   {journal} {Phys. Rev. B}\ }\textbf {\bibinfo {volume} {104}},\ \bibinfo
  {pages} {035124} (\bibinfo {year} {2021})}\BibitemShut {NoStop}%
\bibitem [{\citenamefont {Karlsson}\ \emph {et~al.}(2021)\citenamefont
  {Karlsson}, \citenamefont {van Leeuwen}, \citenamefont {Pavlyukh},
  \citenamefont {Perfetto},\ and\ \citenamefont
  {Stefanucci}}]{karlsson_fast_2021}%
  \BibitemOpen
  \bibfield  {author} {\bibinfo {author} {\bibfnamefont {D.}~\bibnamefont
  {Karlsson}}, \bibinfo {author} {\bibfnamefont {R.}~\bibnamefont {van
  Leeuwen}}, \bibinfo {author} {\bibfnamefont {Y.}~\bibnamefont {Pavlyukh}},
  \bibinfo {author} {\bibfnamefont {E.}~\bibnamefont {Perfetto}},\ and\
  \bibinfo {author} {\bibfnamefont {G.}~\bibnamefont {Stefanucci}},\ }\bibfield
   {title} {\bibinfo {title} {Fast {Green}'s {Function} {Method} for
  {Ultrafast} {Electron}-{Boson} {Dynamics}},\ }\href
  {https://doi.org/10.1103/PhysRevLett.127.036402} {\bibfield  {journal}
  {\bibinfo  {journal} {Phys. Rev. Lett.}\ }\textbf {\bibinfo {volume} {127}},\
  \bibinfo {pages} {036402} (\bibinfo {year} {2021})}\BibitemShut {NoStop}%
\bibitem [{\citenamefont {Pavlyukh}\ \emph
  {et~al.}(2022{\natexlab{a}})\citenamefont {Pavlyukh}, \citenamefont
  {Perfetto}, \citenamefont {Karlsson}, \citenamefont {van Leeuwen},\ and\
  \citenamefont {Stefanucci}}]{pavlyukh_time-linear_2022-1}%
  \BibitemOpen
  \bibfield  {author} {\bibinfo {author} {\bibfnamefont {Y.}~\bibnamefont
  {Pavlyukh}}, \bibinfo {author} {\bibfnamefont {E.}~\bibnamefont {Perfetto}},
  \bibinfo {author} {\bibfnamefont {D.}~\bibnamefont {Karlsson}}, \bibinfo
  {author} {\bibfnamefont {R.}~\bibnamefont {van Leeuwen}},\ and\ \bibinfo
  {author} {\bibfnamefont {G.}~\bibnamefont {Stefanucci}},\ }\bibfield  {title}
  {\bibinfo {title} {Time-linear scaling nonequilibrium {Green}'s function
  methods for real-time simulations of interacting electrons and bosons. {I}.
  {Formalism}},\ }\href {https://doi.org/10.1103/PhysRevB.105.125134}
  {\bibfield  {journal} {\bibinfo  {journal} {Phys. Rev. B}\ }\textbf {\bibinfo
  {volume} {105}},\ \bibinfo {pages} {125134} (\bibinfo {year}
  {2022}{\natexlab{a}})}\BibitemShut {NoStop}%
\bibitem [{\citenamefont {Tuovinen}\ and\ \citenamefont
  {Pavlyukh}(2024)}]{tuovinen_electroluminescence_2024}%
  \BibitemOpen
  \bibfield  {author} {\bibinfo {author} {\bibfnamefont {R.}~\bibnamefont
  {Tuovinen}}\ and\ \bibinfo {author} {\bibfnamefont {Y.}~\bibnamefont
  {Pavlyukh}},\ }\bibfield  {title} {\bibinfo {title} {Electroluminescence
  {Rectification} and {High} {Harmonic} {Generation} in {Molecular}
  {Junctions}},\ }\href {https://doi.org/10.1021/acs.nanolett.4c02609}
  {\bibfield  {journal} {\bibinfo  {journal} {Nano Lett.}\ ,\ \bibinfo {pages}
  {acs.nanolett.4c02609}} (\bibinfo {year} {2024})}\BibitemShut {NoStop}%
\bibitem [{\citenamefont {Joost}\ \emph {et~al.}(2022)\citenamefont {Joost},
  \citenamefont {Schl\"{u}nzen}, \citenamefont {Ohldag}, \citenamefont
  {Bonitz}, \citenamefont {Lackner},\ and\ \citenamefont
  {B\v{r}ezinov\'{a}}}]{joost_dynamically_2022}%
  \BibitemOpen
  \bibfield  {author} {\bibinfo {author} {\bibfnamefont {J.-P.}\ \bibnamefont
  {Joost}}, \bibinfo {author} {\bibfnamefont {N.}~\bibnamefont
  {Schl\"{u}nzen}}, \bibinfo {author} {\bibfnamefont {H.}~\bibnamefont
  {Ohldag}}, \bibinfo {author} {\bibfnamefont {M.}~\bibnamefont {Bonitz}},
  \bibinfo {author} {\bibfnamefont {F.}~\bibnamefont {Lackner}},\ and\ \bibinfo
  {author} {\bibfnamefont {I.}~\bibnamefont {B\v{r}ezinov\'{a}}},\ }\bibfield
  {title} {\bibinfo {title} {Dynamically screened ladder approximation:
  {Simultaneous} treatment of strong electronic correlations and dynamical
  screening out of equilibrium},\ }\href
  {https://doi.org/10.1103/PhysRevB.105.165155} {\bibfield  {journal} {\bibinfo
   {journal} {Phys. Rev. B}\ }\textbf {\bibinfo {volume} {105}},\ \bibinfo
  {pages} {165155} (\bibinfo {year} {2022})}\BibitemShut {NoStop}%
\bibitem [{\citenamefont {Pavlyukh}\ \emph
  {et~al.}(2022{\natexlab{b}})\citenamefont {Pavlyukh}, \citenamefont
  {Perfetto}, \citenamefont {Karlsson}, \citenamefont {van Leeuwen},\ and\
  \citenamefont {Stefanucci}}]{pavlyukh_time-linear_2022}%
  \BibitemOpen
  \bibfield  {author} {\bibinfo {author} {\bibfnamefont {Y.}~\bibnamefont
  {Pavlyukh}}, \bibinfo {author} {\bibfnamefont {E.}~\bibnamefont {Perfetto}},
  \bibinfo {author} {\bibfnamefont {D.}~\bibnamefont {Karlsson}}, \bibinfo
  {author} {\bibfnamefont {R.}~\bibnamefont {van Leeuwen}},\ and\ \bibinfo
  {author} {\bibfnamefont {G.}~\bibnamefont {Stefanucci}},\ }\bibfield  {title}
  {\bibinfo {title} {Time-linear scaling nonequilibrium {Green}'s function
  method for real-time simulations of interacting electrons and bosons. {II}.
  {Dynamics} of polarons and doublons},\ }\href
  {https://doi.org/10.1103/PhysRevB.105.125135} {\bibfield  {journal} {\bibinfo
   {journal} {Phys. Rev. B}\ }\textbf {\bibinfo {volume} {105}},\ \bibinfo
  {pages} {125135} (\bibinfo {year} {2022}{\natexlab{b}})}\BibitemShut
  {NoStop}%
\bibitem [{\citenamefont {Kalvov\'{a}}\ \emph {et~al.}(2019)\citenamefont
  {Kalvov\'{a}}, \citenamefont {Velický},\ and\ \citenamefont
  {\v{S}pi\v{c}ka}}]{kalvova_beyond_2019}%
  \BibitemOpen
  \bibfield  {author} {\bibinfo {author} {\bibfnamefont {A.}~\bibnamefont
  {Kalvov\'{a}}}, \bibinfo {author} {\bibfnamefont {B.}~\bibnamefont
  {Velický}},\ and\ \bibinfo {author} {\bibfnamefont {V.}~\bibnamefont
  {\v{S}pi\v{c}ka}},\ }\bibfield  {title} {\bibinfo {title} {Beyond the
  {Generalized} {Kadanoff}-{Baym} {Ansatz}},\ }\href
  {https://doi.org/10.1002/pssb.201800594} {\bibfield  {journal} {\bibinfo
  {journal} {Phys. Status Solidi B}\ }\textbf {\bibinfo {volume} {256}},\
  \bibinfo {pages} {1800594} (\bibinfo {year} {2019})}\BibitemShut {NoStop}%
\bibitem [{\citenamefont {Kalvov\'{a}}\ \emph {et~al.}(2023)\citenamefont
  {Kalvov\'{a}}, \citenamefont {\v{S}pi\v{c}ka}, \citenamefont {Velický},\
  and\ \citenamefont {Lipavský}}]{kalvova_dynamical_2023}%
  \BibitemOpen
  \bibfield  {author} {\bibinfo {author} {\bibfnamefont {A.}~\bibnamefont
  {Kalvov\'{a}}}, \bibinfo {author} {\bibfnamefont {V.}~\bibnamefont
  {\v{S}pi\v{c}ka}}, \bibinfo {author} {\bibfnamefont {B.}~\bibnamefont
  {Velický}},\ and\ \bibinfo {author} {\bibfnamefont {P.}~\bibnamefont
  {Lipavský}},\ }\bibfield  {title} {\bibinfo {title} {Dynamical vertex
  correction to the generalized {Kadanoff}-{Baym} {Ansatz}},\ }\href
  {https://doi.org/10.1209/0295-5075/acad9b} {\bibfield  {journal} {\bibinfo
  {journal} {Europhys. Lett.}\ }\textbf {\bibinfo {volume} {141}},\ \bibinfo
  {pages} {16002} (\bibinfo {year} {2023})}\BibitemShut {NoStop}%
\bibitem [{\citenamefont {Kalvov\'{a}}\ \emph {et~al.}(2024)\citenamefont
  {Kalvov\'{a}}, \citenamefont {\v{S}pi\v{c}ka}, \citenamefont {Velický},\
  and\ \citenamefont {Lipavský}}]{kalvova_fast_2024}%
  \BibitemOpen
  \bibfield  {author} {\bibinfo {author} {\bibfnamefont {A.}~\bibnamefont
  {Kalvov\'{a}}}, \bibinfo {author} {\bibfnamefont {V.}~\bibnamefont
  {\v{S}pi\v{c}ka}}, \bibinfo {author} {\bibfnamefont {B.}~\bibnamefont
  {Velický}},\ and\ \bibinfo {author} {\bibfnamefont {P.}~\bibnamefont
  {Lipavský}},\ }\bibfield  {title} {\bibinfo {title} {Fast corrections to the
  generalized {Kadanoff}-{Baym} ansatz},\ }\href
  {https://doi.org/10.1103/PhysRevB.109.134306} {\bibfield  {journal} {\bibinfo
   {journal} {Phys. Rev. B}\ }\textbf {\bibinfo {volume} {109}},\ \bibinfo
  {pages} {134306} (\bibinfo {year} {2024})}\BibitemShut {NoStop}%
\bibitem [{\citenamefont {Stefanucci}\ and\ \citenamefont
  {Perfetto}(2024)}]{stefanucci_semiconductor_2024}%
  \BibitemOpen
  \bibfield  {author} {\bibinfo {author} {\bibfnamefont {G.}~\bibnamefont
  {Stefanucci}}\ and\ \bibinfo {author} {\bibfnamefont {E.}~\bibnamefont
  {Perfetto}},\ }\bibfield  {title} {\bibinfo {title} {Semiconductor
  electron-phonon equations: {A} rung above {Boltzmann} in the many-body
  ladder},\ }\href {https://doi.org/10.21468/SciPostPhys.16.3.073} {\bibfield
  {journal} {\bibinfo  {journal} {SciPost Physics}\ }\textbf {\bibinfo {volume}
  {16}},\ \bibinfo {pages} {073} (\bibinfo {year} {2024})}\BibitemShut
  {NoStop}%
\bibitem [{\citenamefont {Hu}\ \emph {et~al.}(2010)\citenamefont {Hu},
  \citenamefont {Xu},\ and\ \citenamefont {Yan}}]{hu_communication:_2010}%
  \BibitemOpen
  \bibfield  {author} {\bibinfo {author} {\bibfnamefont {J.}~\bibnamefont
  {Hu}}, \bibinfo {author} {\bibfnamefont {R.-X.}\ \bibnamefont {Xu}},\ and\
  \bibinfo {author} {\bibfnamefont {Y.}~\bibnamefont {Yan}},\ }\bibfield
  {title} {\bibinfo {title} {Communication: {Pad\'{e}} spectrum decomposition
  of {Fermi} function and {Bose} function},\ }\href
  {https://doi.org/10.1063/1.3484491} {\bibfield  {journal} {\bibinfo
  {journal} {J. Chem. Phys.}\ }\textbf {\bibinfo {volume} {133}},\ \bibinfo
  {pages} {101106} (\bibinfo {year} {2010})}\BibitemShut {NoStop}%
\bibitem [{sup()}]{suppmat}%
  \BibitemOpen
  \href@noop {} {}\bibinfo {note} {See Supplemental Material, which includes
  Ref.~\cite{pavlyukh_cheers_2023} for more detailed discussion}\BibitemShut
  {NoStop}%
\bibitem [{com()}]{comment}%
  \BibitemOpen
  \href@noop {} {}\bibinfo {note} {Note that bias is not incorporated in the
  definition of the tunneling rate, Eq.~\eqref{eq:lorentzian}, and, therefore,
  must be included in the arguments of $\Gamma_\a$ and $\Gamma_\b$ in the
  Landauer–B\"{u}ttiker formula.}\BibitemShut {Stop}%
\bibitem [{\citenamefont {Schl\"{u}nzen}\ and\ \citenamefont
  {Bonitz}(2016)}]{schlunzen_nonequilibrium_2016}%
  \BibitemOpen
  \bibfield  {author} {\bibinfo {author} {\bibfnamefont {N.}~\bibnamefont
  {Schl\"{u}nzen}}\ and\ \bibinfo {author} {\bibfnamefont {M.}~\bibnamefont
  {Bonitz}},\ }\bibfield  {title} {\bibinfo {title} {Nonequilibrium {Green}
  {Functions} {Approach} to {Strongly} {Correlated} {Fermions} in {Lattice}
  {Systems}},\ }\href {https://doi.org/10.1002/ctpp.201610003} {\bibfield
  {journal} {\bibinfo  {journal} {Contributions to Plasma Physics}\ }\textbf
  {\bibinfo {volume} {56}},\ \bibinfo {pages} {5} (\bibinfo {year}
  {2016})}\BibitemShut {NoStop}%
\bibitem [{\citenamefont {Tuovinen}\ \emph {et~al.}(2020)\citenamefont
  {Tuovinen}, \citenamefont {Gole\v{z}}, \citenamefont {Eckstein},\ and\
  \citenamefont {Sentef}}]{tuovinen_comparing_2020}%
  \BibitemOpen
  \bibfield  {author} {\bibinfo {author} {\bibfnamefont {R.}~\bibnamefont
  {Tuovinen}}, \bibinfo {author} {\bibfnamefont {D.}~\bibnamefont {Gole\v{z}}},
  \bibinfo {author} {\bibfnamefont {M.}~\bibnamefont {Eckstein}},\ and\
  \bibinfo {author} {\bibfnamefont {M.~A.}\ \bibnamefont {Sentef}},\ }\bibfield
   {title} {\bibinfo {title} {Comparing the generalized {Kadanoff}-{Baym}
  ansatz with the full {Kadanoff}-{Baym} equations for an excitonic insulator
  out of equilibrium},\ }\href {https://doi.org/10.1103/PhysRevB.102.115157}
  {\bibfield  {journal} {\bibinfo  {journal} {Phys. Rev. B}\ }\textbf {\bibinfo
  {volume} {102}},\ \bibinfo {pages} {115157} (\bibinfo {year}
  {2020})}\BibitemShut {NoStop}%
\bibitem [{\citenamefont {Tuovinen}(2021)}]{tuovinen_electron_2021}%
  \BibitemOpen
  \bibfield  {author} {\bibinfo {author} {\bibfnamefont {R.}~\bibnamefont
  {Tuovinen}},\ }\bibfield  {title} {\bibinfo {title} {Electron correlation
  effects in superconducting nanowires in and out of equilibrium},\ }\href
  {https://doi.org/10.1088/1367-2630/ac1898} {\bibfield  {journal} {\bibinfo
  {journal} {New J. Phys.}\ }\textbf {\bibinfo {volume} {23}},\ \bibinfo
  {pages} {083024} (\bibinfo {year} {2021})}\BibitemShut {NoStop}%
\bibitem [{\citenamefont {Stahl}\ \emph {et~al.}(2022)\citenamefont {Stahl},
  \citenamefont {Dasari}, \citenamefont {Li}, \citenamefont {Picano},
  \citenamefont {Werner},\ and\ \citenamefont {Eckstein}}]{stahl_memory_2022}%
  \BibitemOpen
  \bibfield  {author} {\bibinfo {author} {\bibfnamefont {C.}~\bibnamefont
  {Stahl}}, \bibinfo {author} {\bibfnamefont {N.}~\bibnamefont {Dasari}},
  \bibinfo {author} {\bibfnamefont {J.}~\bibnamefont {Li}}, \bibinfo {author}
  {\bibfnamefont {A.}~\bibnamefont {Picano}}, \bibinfo {author} {\bibfnamefont
  {P.}~\bibnamefont {Werner}},\ and\ \bibinfo {author} {\bibfnamefont
  {M.}~\bibnamefont {Eckstein}},\ }\bibfield  {title} {\bibinfo {title} {Memory
  truncated {Kadanoff}-{Baym} equations},\ }\href
  {https://doi.org/10.1103/PhysRevB.105.115146} {\bibfield  {journal} {\bibinfo
   {journal} {Phys. Rev. B}\ }\textbf {\bibinfo {volume} {105}},\ \bibinfo
  {pages} {115146} (\bibinfo {year} {2022})}\BibitemShut {NoStop}%
\bibitem [{\citenamefont {Makait}\ \emph {et~al.}(2023)\citenamefont {Makait},
  \citenamefont {Fajardo},\ and\ \citenamefont
  {Bonitz}}]{makait_time-dependent_2023}%
  \BibitemOpen
  \bibfield  {author} {\bibinfo {author} {\bibfnamefont {C.}~\bibnamefont
  {Makait}}, \bibinfo {author} {\bibfnamefont {F.~B.}\ \bibnamefont
  {Fajardo}},\ and\ \bibinfo {author} {\bibfnamefont {M.}~\bibnamefont
  {Bonitz}},\ }\bibfield  {title} {\bibinfo {title} {Time-dependent charged
  particle stopping in quantum plasmas: testing the {G1}-{G2} scheme for
  quasi-one-dimensional systems},\ }\href
  {https://doi.org/10.1002/ctpp.202300008} {\bibfield  {journal} {\bibinfo
  {journal} {Contributions to Plasma Physics}\ }\textbf {\bibinfo {volume}
  {63}},\ \bibinfo {pages} {e202300008} (\bibinfo {year} {2023})}\BibitemShut
  {NoStop}%
\bibitem [{\citenamefont {Reeves}\ \emph {et~al.}(2023)\citenamefont {Reeves},
  \citenamefont {Zhu}, \citenamefont {Yang},\ and\ \citenamefont
  {Vl\v{c}ek}}]{reeves_unimportance_2023}%
  \BibitemOpen
  \bibfield  {author} {\bibinfo {author} {\bibfnamefont {C.~C.}\ \bibnamefont
  {Reeves}}, \bibinfo {author} {\bibfnamefont {Y.}~\bibnamefont {Zhu}},
  \bibinfo {author} {\bibfnamefont {C.}~\bibnamefont {Yang}},\ and\ \bibinfo
  {author} {\bibfnamefont {V.}~\bibnamefont {Vl\v{c}ek}},\ }\bibfield  {title}
  {\bibinfo {title} {Unimportance of memory for the time nonlocal components of
  the {Kadanoff}-{Baym} equations},\ }\href
  {https://doi.org/10.1103/PhysRevB.108.115152} {\bibfield  {journal} {\bibinfo
   {journal} {Phys. Rev. B}\ }\textbf {\bibinfo {volume} {108}},\ \bibinfo
  {pages} {115152} (\bibinfo {year} {2023})}\BibitemShut {NoStop}%
\bibitem [{\citenamefont {Strasberg}\ \emph {et~al.}(2017)\citenamefont
  {Strasberg}, \citenamefont {Schaller}, \citenamefont {Brandes},\ and\
  \citenamefont {Esposito}}]{strasberg_quantum_2017}%
  \BibitemOpen
  \bibfield  {author} {\bibinfo {author} {\bibfnamefont {P.}~\bibnamefont
  {Strasberg}}, \bibinfo {author} {\bibfnamefont {G.}~\bibnamefont {Schaller}},
  \bibinfo {author} {\bibfnamefont {T.}~\bibnamefont {Brandes}},\ and\ \bibinfo
  {author} {\bibfnamefont {M.}~\bibnamefont {Esposito}},\ }\bibfield  {title}
  {\bibinfo {title} {Quantum and information thermodynamics: A unifying
  framework based on repeated interactions},\ }\href
  {https://doi.org/10.1103/PhysRevX.7.021003} {\bibfield  {journal} {\bibinfo
  {journal} {Phys. Rev. X}\ }\textbf {\bibinfo {volume} {7}},\ \bibinfo {pages}
  {021003} (\bibinfo {year} {2017})}\BibitemShut {NoStop}%
\bibitem [{\citenamefont {Portugal}\ \emph {et~al.}(2024)\citenamefont
  {Portugal}, \citenamefont {Brange},\ and\ \citenamefont
  {Flindt}}]{portugal_heat_2024}%
  \BibitemOpen
  \bibfield  {author} {\bibinfo {author} {\bibfnamefont {P.}~\bibnamefont
  {Portugal}}, \bibinfo {author} {\bibfnamefont {F.}~\bibnamefont {Brange}},\
  and\ \bibinfo {author} {\bibfnamefont {C.}~\bibnamefont {Flindt}},\
  }\bibfield  {title} {\bibinfo {title} {Heat pulses in electron quantum
  optics},\ }\href {https://doi.org/10.1103/PhysRevLett.132.256301} {\bibfield
  {journal} {\bibinfo  {journal} {Phys. Rev. Lett.}\ }\textbf {\bibinfo
  {volume} {132}},\ \bibinfo {pages} {256301} (\bibinfo {year}
  {2024})}\BibitemShut {NoStop}%
\bibitem [{\citenamefont {Boschini}\ \emph {et~al.}(2024)\citenamefont
  {Boschini}, \citenamefont {Zonno},\ and\ \citenamefont
  {Damascelli}}]{boschini_time-resolved_2024}%
  \BibitemOpen
  \bibfield  {author} {\bibinfo {author} {\bibfnamefont {F.}~\bibnamefont
  {Boschini}}, \bibinfo {author} {\bibfnamefont {M.}~\bibnamefont {Zonno}},\
  and\ \bibinfo {author} {\bibfnamefont {A.}~\bibnamefont {Damascelli}},\
  }\bibfield  {title} {\bibinfo {title} {Time-resolved {ARPES} studies of
  quantum materials},\ }\href {https://doi.org/10.1103/RevModPhys.96.015003}
  {\bibfield  {journal} {\bibinfo  {journal} {Rev. Mod. Phys.}\ }\textbf
  {\bibinfo {volume} {96}},\ \bibinfo {pages} {015003} (\bibinfo {year}
  {2024})}\BibitemShut {NoStop}%
\bibitem [{\citenamefont {Pavlyukh}\ \emph {et~al.}()\citenamefont {Pavlyukh},
  \citenamefont {Tuovinen}, \citenamefont {Perfetto},\ and\ \citenamefont
  {Stefanucci}}]{pavlyukh_cheers_2023}%
  \BibitemOpen
  \bibfield  {author} {\bibinfo {author} {\bibfnamefont {Y.}~\bibnamefont
  {Pavlyukh}}, \bibinfo {author} {\bibfnamefont {R.}~\bibnamefont {Tuovinen}},
  \bibinfo {author} {\bibfnamefont {E.}~\bibnamefont {Perfetto}},\ and\
  \bibinfo {author} {\bibfnamefont {G.}~\bibnamefont {Stefanucci}},\ }\bibfield
   {title} {\bibinfo {title} {Cheers: {A} {Linear}‐{Scaling} {KBE} + {GKBA}
  {Code}},\ }\href {https://doi.org/10.1002/pssb.202300504} {\bibfield
  {journal} {\bibinfo  {journal} {Phys. Status Solidi B}\ }\textbf {\bibinfo
  {volume} {2023}},\ \bibinfo {pages} {2300504}}\BibitemShut {NoStop}%
\end{thebibliography}

\end{document}